\documentclass[english,aps,preprint,nofootinbib]{revtex4}
\usepackage[T1]{fontenc}
\usepackage[latin9]{inputenc}
\usepackage{textcomp}
\usepackage{esint}
\usepackage{color}

\makeatletter

\providecommand{\LyX}{L\kern-.1667em\lower.25em\hbox{Y}\kern-.125emX\@}

\definecolor{BLACK}{gray}{0}
\definecolor{WHITE}{gray}{1}
\definecolor{RED}{rgb}{1,0,0}
\definecolor{GREEN}{rgb}{0,1,0}
\definecolor{BLUE}{rgb}{0,0,1}
\definecolor{CYAN}{cmyk}{1,0,0,0}
\definecolor{MAGENTA}{cmyk}{0,1,0,0}
\definecolor{YELLOW}{cmyk}{0,0,1,0}
\definecolor{lightgray}{rgb}{.7,.7,.7}

\makeatother

\usepackage{babel}

\begin{document}

\preprint{This line only printed with preprint option}

\title{Incompressible Navier-Stokes Equation from Einstein-Maxwell and Gauss-Bonnet-Maxwell
Theories}

\author{Chao Niu and Yu Tian}

\affiliation{College of Physical Sciences, Graduate University of
Chinese Academy of Sciences, Beijing 100049, China}

\author{Xiao-Ning Wu}

\affiliation{Institute of Mathematics, Academy of Mathematics and
System Science, {CAS}, Beijing 100190, China}

\affiliation{Hua Loo-Keng Key Laboratory of Mathematics, {CAS},
Beijing 100190, China}

\author{Yi Ling}

\affiliation{Center for Relativistic Astrophysics and High Energy
Physics, Department of Physics, Nanchang University, 330031, China}

\affiliation{Institute of Mathematics, Academy of Mathematics and
System Science, {CAS}, Beijing 100190, China}

\begin{abstract}
The dual fluid description for a general cutoff surface at radius
$r=r_{c}$ outside the horizon in the charged AdS black brane bulk
space-time is investigated, first in the Einstein-Maxwell theory.
Under the non-relativistic long-wavelength expansion with parameter
$\epsilon$, the coupled Einstein-Maxwell equations are solved up to
$\mathcal{O}(\epsilon^{2})$. The incompressible Navier-Stokes
equation with external force density is obtained as the constraint
equation at the cutoff surface. For non-extremal black brane, the
viscosity of the dual fluid is determined by the regularity of the
metric fluctuation at the horizon, whose ratio to entropy density
$\eta/s$ is independent of both the cutoff $r_{c}$ and the
black brane charge. Then, we extend our discussion to the
Gauss-Bonnet-Maxwell case, where the incompressible Navier-Stokes
equation with external force density is also obtained at a general
cutoff surface. In this case, it turns out that the ratio $\eta/s$
is independent of the cutoff $r_{c}$ but dependent on the charge
density of the black brane.
\end{abstract}
\maketitle

\section{Introduction}

It has long been known that excitations of a black hole horizon
dissipate like those of a fluid with viscosity $\eta=\frac{1}{16\pi
G}$ \cite{Damour2} (see also \cite{Damour3,P-T,TPM,D-L}), which
together with the Bekenstein-Hawking entropy density
$s=\frac{1}{4G}$ yields the dimensionless ratio
$\frac{\eta}{s}=\frac{1}{4\pi}$. Interestingly, this fact has been
related to similar results \cite{PSS} in the AdS/CFT correspondence,
which regards the change of radius $r$ as equivalent to the
renormalization group (RG) flow
\cite{KSS,S-S,B-L,I-L,EFO,E-O,Paulos,B-D,ABD2} and the case of black
hole horizon corresponds to the IR limit of this flow. Especially,
it is proved that the ratio $\frac{\eta}{s}$ does not run with the
RG flow \cite{I-L,Strominger}, and so the universality of this ratio
in both the horizon fluid and the standard AdS/CFT follows.

According to the general holographic dictionary \cite{Witten}, the
Brown-York tensor of the bulk gravity is dual to the expectation value
of the stress-energy tensor of the boundary field theory \cite{Myers}.
Under the long-wavelength limit, the boundary theory can be described
by hydrodynamics, known as the gravity/fluid duality \cite{BHMR,Rangamani}.
Especially, under certain non-relativistic limit (or scaling), the
boundary hydrodynamics takes the form of the standard incompressible
Navier-Stokes equation \cite{BMW,EFO}.
Recently, Bredberg \textit{et al} give the precise definition of the
boundary theory on an arbitrary cutoff surface $r=r_{c}$ outside
the horizon in the Rindler bulk space-time, which reduces the bulk
gravitational dynamics to the incompressible Navier-Stokes equation
on the boundary \cite{Strominger2}. This framework is extended to
the AdS black brane case {\cite{BCLR}}, as well as in the Gauss-Bonnet gravity \cite{Cai}.

In this Letter, we consider an arbitrary cutoff surface $r=r_{c}$
outside the horizon in the charged AdS black brane bulk space-times
in both the Einstein-Maxwell and Gauss-Bonnet-Maxwell theories,
further extending the framework of \cite{Strominger2} and
\cite{Cai}. Under the non-relativistic long-wavelength expansion
with parameter $\epsilon$, the coupled Einstein-Maxwell (or
Gauss-Bonnet-Maxwell) equations are solved up to
$\mathcal{O}(\epsilon^{2})$. The incompressible Navier-Stokes
equation with external force density is obtained as the constraint
equation at the cutoff surface, with a viscosity $\eta$ satisfying
the universality relation $\frac{\eta}{s}=\frac{1}{4\pi}$
independent of both the cutoff $r_{c}$ and the black brane charge in
the Einstein-Maxwell case. In the Gauss-Bonnet-Maxwell case, the
incompressible Navier-Stokes equation with external force density is
as well obtained at a general cutoff surface. However, it turns out
that the regularity condition at the horizon determines the ratio
$\frac{\eta}{s}=\frac{1}{4\pi}\{1-2(n-4){\alpha}[n-1-(n-3) q_h^2]\}$
with ${\alpha}$ the Gauss-Bonnet coupling constant and $n$ the
space-time dimensionality, which is independent of the cutoff
$r_{c}$ but dependent on the charge density $q_{h}$ of the black
brane.

The rest of the Letter is organized as follows. In Sec.~II, we
present the metric and electromagnetic background and introduce the
non-relativistic long-wavelength expansion, focusing on the
Einstein-Maxwell case, where the bulk equations of motion are solved
up to $\mathcal{O}(\epsilon^{2})$. In Sec.~III, the dual fluid on
the cutoff surface is analyzed with the incompressible Navier-Stokes
equation obtained in the Einstein-Maxwell case. In Sec.~IV, we
extend the above discussion to the Gauss-Bonnet-Maxwell case.

\section{Metric and electromagnetic configurations}

We consider the standard $n$-dimensional Einstein-Maxwell gravity
with the action\[
I=\frac{1}{16\pi G}\int d^{n}x\sqrt{-g}(R-2\Lambda)-\frac{1}{4}\int d^{n}x\sqrt{-g}F_{\mu\nu}F^{\mu\nu},\]
where $\Lambda=-\frac{(n-2)(n-1)}{2l^{2}}$ is the negative cosmological
constant. The corresponding equations of motion are\begin{eqnarray}
G_{\mu\nu}+\Lambda g_{\mu\nu}+8\pi GT_{\mu\nu} & = & 0,\nonumber \\
\nabla_{\mu}F^{\mu\nu} & = & 0,\label{eq:E-M}\end{eqnarray}
where $T_{\mu\nu}=\frac{1}{4}g_{\mu\nu}F_{\rho\sigma}F^{\rho\sigma}-F_{\mu\rho}F_{\nu}^{\ \rho}$
is the stress-energy tensor of the electromagnetic field.

Our background is the charged black brane solution\begin{eqnarray}
ds^{2} & = & -f(r)dt^{2}+2drdt+r^{2}dx^{a}dx^{a},\qquad f(r)=\frac{r^{2}}{l^{2}}-\frac{2m}{r^{n-3}}+\frac{Q^{2}}{r^{2n-6}},\nonumber \\
A & = & \sqrt{\frac{n-2}{8\pi(n-3)G}}\frac{Q}{r^{n-3}}dt\label{eq:background}\end{eqnarray}
under Eddington-Finkelstein coordinates, where the index $a$ runs
from $2$ to $n-1$, $m$ is the mass parameter and $Q$ the charge
parameter. The corresponding electromagnetic field strength is\[
F=\sqrt{\frac{(n-2)(n-3)}{8\pi G}}\frac{Q}{r^{n-2}}dt\wedge dr.\]
For convenience, we take the AdS radius $l=1$ hereafter.

The induced metric on the cutoff surface $r=r_{c}$ outside the horizon
is\begin{equation}
ds_{c}^{2}=-f(r_{c})dt^{2}+r_{c}^{2}dx^{a}dx^{a},\label{eq:cutoff}\end{equation}
which is flat and kept fixed when perturbing the bulk metric. In order
to introduce the fluid degrees of freedom $v^{a}$ (the velocity)
and $P$ (the pressure), two types of diffeomorphisms that keep (\ref{eq:cutoff})
invariant are taken:
\begin{enumerate}
\item Lorentz boost\[
\left\{ \begin{array}{c}
\sqrt{f(r_{c})}t\to\gamma(\sqrt{f(r_{c})}t-\beta_{a}r_{c}x^{a}),\qquad\gamma\equiv1/\sqrt{1-\beta^{2}},\\
r_{c}x^{a}\to(\delta_{b}^{a}-\frac{\beta^{a}\beta_{b}}{\beta^{2}})r_{c}x^{b}+\gamma(\frac{\beta^{a}\beta_{b}}{\beta^{2}}r_{c}x^{b}-\beta^{a}\sqrt{f(r_{c})}t)\end{array}\right.\]
with $\beta^{a}\equiv\frac{r_{c}}{\sqrt{f(r_{c})}}v^{a}$;
\item Special rescaling\begin{equation}
r\to(1-P)r,\qquad t\to\sqrt{\frac{f(r_{c})}{f[(1-P)r_{c}]}}t,\qquad x^{a}\to\frac{r_{c}}{(1-P)r_{c}}x^{a}\label{eq:special}\end{equation}
of $r$, $t$ and $x^{a}$.
\end{enumerate}
Then we promote $v^{a}$ and $P$ to be the velocity field
$v^{a}(t,x)$ and the pressure field $P(t,x)$, which makes the
transformed bulk metric and electromagnetic field no longer be
solution of the Einstein-Maxwell equations (\ref{eq:E-M}). We also
adopt the non-relativistic long-wavelength expansion parameterized
by $\epsilon\to0$ and the scaling \cite{Strominger2}\begin{equation}
\partial_{t} \sim
\epsilon^{2},\qquad\partial_{a}\sim\epsilon,\qquad\partial_{r}\sim1,\qquad
P \sim \epsilon^{2},\qquad
v^{a}\sim\epsilon\label{eq:scaling}\end{equation} with
$\partial_{a}\equiv\frac{\partial}{\partial x^{a}}$, under which the
perturbed bulk Einstein-Maxwell equations can be solved order by
order. Up to $\mathcal{O}(\epsilon^{2})$, the transformed bulk
metric by both types of diffeomorphisms has been given in \cite{Cai}
(actually for arbitrary $f(r)$) as%
\footnote{Note that our $f(r)$ corresponds to $r^{2}f(r)$ in \cite{Cai},
and we do not distinguish $v_{a}$ from $v^{a}$.%
}\begin{eqnarray}
ds^{2} & = & -f(r)dt^{2}+2drdt+r^{2}dx^{a}dx^{a}-2r^{2}(1-\frac{r_{c}^{2}f(r)}{r^{2}f(r_{c})})v^{a}dx^{a}dt-2\frac{r_{c}^{2}v^{a}}{f(r_{c})}dx^{a}dr\nonumber \\
 &  & +r^{2}(1-\frac{r_{c}^{2}f(r)}{r^{2}f(r_{c})})(v^{2}dt^{2}+\frac{r_{c}^{2}v^{a}v^{b}}{f(r_{c})}dx^{a}dx^{b})+\frac{r_{c}^{2}v^{2}}{f(r_{c})}drdt\nonumber \\
 &  & +f(r)(\frac{rf^{\prime}(r)}{f(r)}-\frac{r_{c}f^{\prime}(r_{c})}{f(r_{c})})Pdt^{2}+(\frac{r_{c}f^{\prime}(r_{c})}{f(r_{c})}-2)Pdrdt+\mathcal{O}(\epsilon^{3}).\label{eq:metric}\end{eqnarray}

The bulk electromagnetic field should also be transformed by the
above two types of diffeomorphisms. After promoting $v^{a}$ and $P$
to be $(t,x)$-dependent (but $r$-independent) fields and adopting
the scaling (\ref{eq:scaling}), the perturbed electromagnetic field
can be straightforwardly worked out as\begin{equation}
A=\sqrt{\frac{n-2}{8\pi(n-3)G}}\frac{Q}{r^{n-3}}[dt-\frac{r_{c}^{2}}{f(r_{c})}v^{a}dx^{a}+\frac{r_{c}^{2}}{2f(r_{c})}v^{2}dt+(n-3)Pdt+\frac{f^{\prime}(r_{c})}{2f(r_{c})}r_{c}Pdt]+\mathcal{O}(\epsilon^{3}),\label{eq:em_field}\end{equation}
up to $\mathcal{O}(\epsilon^{2})$. It should be noted that in this charged configuration, there are
gravitational degrees of freedom (DoF) and electromagnetic DoF. If the electromagnetic DoF are {taken into account},
there is a boundary current dual to the bulk electromagnetic field (see e.g. \cite{Strominger}). In
our approach, we only consider the DoF induced by the above two kinds of
(lifted) diffeomorphisms,
which can be roughly regarded as gravitational, and do not turn on
the independent electromagnetic DoF. That is, our focus is on the
influence of the charged AdS black brane, as the background
(compared to the uncharged one), on the properties of the dual
fluid, especially the Navier-Stokes equation and the viscosity $\eta$ (or the ratio
$\frac{\eta}{s}$) appearing there. It turns out that the resulting dual
charge density is constant at visible orders, and so the conservation law of the boundary current
just coincides with the incompressibility condition of the dual fluid
(see (9) below) and does not give any new equation.\footnote{The independent
electromagnetic DoF can be turned on, e.g. by lifting the black brane
charge $Q$ to a function of $(t,x)$, which will result in a non-constant dual charge density
at visible orders.}

Now we should substitute the
perturbed metric (\ref{eq:metric}) and electromagnetic field
(\ref{eq:em_field}) into the Einstein-Maxwell equations
(\ref{eq:E-M}) and see if they have already solved the equations up
to $\mathcal{O}(\epsilon^{2})$. Different from the asymptotically
flat case \cite{Skenderis}, the perturbed metric (together with the
perturbed electromagnetic field) only solves the Einstein(-Maxwell)
equations (\ref{eq:E-M}) up to $\mathcal{O}(\epsilon)$, even if we
imposed the $\mathcal{O}(\epsilon^{2})$ constraint equation
$\partial_{a}v^{a}=0$ (incompressibility of the dual fluid on the
cutoff surface). It is known in the case without electromagnetic
field that a correction term
\begin{equation}\label{eq:correction}
r^{2}\mathcal{F}(r)(\partial_{a}v^{b}+\partial_{b}v^{a})dx^{a}dx^{b}
\end{equation}
at $\mathcal{O}(\epsilon^{2})$ should be added to the metric
(\ref{eq:metric}) \cite{Cai}. It turns out that this prescription
also applies to our case, without the need of additional correction
terms for the perturbed electromagnetic field (\ref{eq:em_field}).
In fact, the Maxwell equations in (\ref{eq:E-M}) are automatically
satisfied at $\mathcal{O}(1)$ and $\mathcal{O}(\epsilon)$, while at
$\mathcal{O}(\epsilon^{2})$ lead to two equations\begin{equation}
\partial_{a}v^{a} = 0,\label{eq:incompress}\qquad
\mathcal{F}^{\prime}(r)\partial_{a}v^{a} = 0,
\end{equation} which can be solved altogether by the
incompressibility condition. Then from the Einstein equations in
(\ref{eq:E-M}) at $\mathcal{O}(\epsilon^{2})$ there is essentially
only one requirement\[
r^{n-2}f(r)\mathcal{F}^{\prime\prime}(r)+r^{n-3}[(n-2)f(r)+rf^{\prime}(r)]\mathcal{F}^{\prime}(r)+(n-2)r^{n-3}=0\]
for $\mathcal{F}(r)$ to satisfy,%
\footnote{We have explicitly obtained the above form of equation for $4\le n\le10$.%
} which can be solved as\[
\mathcal{F}^{\prime}(r)=-(1-\frac{C}{r^{n-2}})\frac{1}{f(r)}.\]
Here the integration constant $C$ can be determined as $C=r_{h}^{n-2}$
with $r_{h}$ the horizon radius by the regularity condition of the
perturbed metric at the horizon, provided our charged black brane
is non-extremal and so $f(r)$ only has a simple zero at $r=r_{h}$.
The explicit form of $\mathcal{F}(r)$ is irrelevant to the succeeding
discussions, where the one more integration constant can be determined
such that the induced metric (\ref{eq:cutoff}) is kept
fixed to all orders in $\epsilon$.

\section{Dual fluid on the cutoff surface}

On an arbitrary cutoff surface $r=r_{c}$ outside the horizon, there
is {firstly a} thermodynamic description of the (equilibrium)
fluid dual to the background configuration (\ref{eq:background}). In
fact, the Brown-York tensor \cite{B-Y}\begin{equation}
t_{ij}=\frac{1}{8\pi
G}(Kg_{ij}-K_{ij}-\mathcal{C}g_{ij})\label{eq:B-Y}\end{equation} on
the cutoff surface, with $K_{ij}$ its extrinsic curvature and
$K\equiv g^{ij}K_{ij}$, is
\begin{equation} t_{ij}dx^{i}dx^{j} = \frac{1}{8\pi
G}[-\sqrt{f(r_{c})}\frac{(n-2)f(r_{c})}{r_{c}}dt^{2}+\frac{r_{c}^{2}}{\sqrt{f(r_{c})}}(\frac{f^{\prime}(r_{c})}{2}+\frac{(n-3)f(r_{c})}{r_{c}})dx^{a}dx^{a}
-\mathcal{C}ds_{c}^{2}],\label{eq:B-Y_c}\end{equation}
which is identified with the stress-energy tensor of the dual fluid \cite{Strominger}. On the other
hand, the stress-energy tensor of a (relativistic) fluid in
equilibrium is\begin{equation}
t_{ij}=(\rho+p)u_{i}u_{j}+pg_{ij}\label{eq:fluid}\end{equation} with
$\rho$ the energy density, $p$ the pressure and
$u^{i}=\frac{1}{\sqrt{-g_{tt}}}(1,0,\cdots,0)$ the normalized fluid
four-velocity. Inclusion of the constant
$\mathcal{C}$ in (\ref{eq:B-Y}) is equivalent to the replacement\[
p\to p-\frac{\mathcal{C}}{8\pi
G},\qquad\rho\to\rho+\frac{\mathcal{C}}{8\pi G},\] which leaves the
combination $\rho+p$ invariant, so we can omit $\mathcal{C}$
if we only consider this combination.%
\footnote{Note that $\mathcal{C}$ is essential for the regularity of $t_{ij}$
as $r_{c}\to\infty$ \cite{B-K}.%
} Comparing (\ref{eq:B-Y_c}) and (\ref{eq:fluid}), we
find\begin{equation} \rho+p=\frac{1}{8\pi
G\sqrt{f(r_{c})}}(\frac{f^{\prime}(r_{c})}{2}-\frac{f(r_{c})}{r_{c}})=\frac{r_{c}^{2}}{16\pi
G\sqrt{f(r_{c})}}(\frac{f(r)}{r^{2}})_{c}^{\prime}.\label{eq:invariant}\end{equation}
Noticing the entropy density\begin{equation}
s_{c}=\frac{1}{4G}\frac{r_{h}^{n-2}}{r_{c}^{n-2}}\label{eq:entropy}\end{equation}
on the cutoff surface, the local Hawking temperature
\begin{equation}\label{eq:temperature}
T_{c}=\frac{1}{\sqrt{f(r_{c})}}\frac{f^{\prime}(r_{h})}{4\pi},
\end{equation}
the explicit form of $f(r)$ in (\ref{eq:background}) and the horizon
condition $f(r_{h})=0$, we have the familiar thermodynamic
relation\begin{eqnarray} \rho+p-s_{c}T_{c} & = &
q_{c}\mu_{c}\label{eq:Gibbs}\end{eqnarray} with (up to some
unimportant constant factor) the charge density
\begin{equation}\label{eq:charge}
q_{c}=\frac{Q}{r_{c}^{n-2}}
\end{equation}
and the corresponding chemical potential
\begin{equation}\label{eq:chemical}
\mu_{c}=\frac{n-2}{8\pi
G\sqrt{f(r_{c})}}(\frac{Q}{r_{h}^{n-3}}-\frac{Q}{r_{c}^{n-3}}),
\end{equation}
which is just a (red-shifted) electric potential difference.

In the perturbed case (\ref{eq:metric}), the Brown-York tensor can be worked out as\begin{eqnarray}
8\pi Gt_{ij}dx^{i}dx^{j} & = & -\sqrt{f(r_{c})}\frac{(n-2)f(r_{c})}{r_{c}}dt^{2}+\frac{r_{c}^{2}}{\sqrt{f(r_{c})}}(\frac{f^{\prime}(r_{c})}{2}+\frac{(n-3)f(r_{c})}{r_{c}})dx^{a}dx^{a}-\mathcal{C}ds_{c}^{2}\nonumber \\
 &  & -\frac{r_{c}^{4}}{\sqrt{f(r_{c})}}(\frac{f(r)}{r^{2}})_{c}^{\prime}v^{a}dx^{a}dt\nonumber \\
 &  & +\frac{r_{c}^{2}}{2\sqrt{f(r_{c})}}[(n-2)f(r_{c})P+r_{c}^{2}v^{2}](\frac{f(r)}{r^{2}})_{c}^{\prime}dt^{2}+\frac{r_{c}^{6}}{2\sqrt{f(r_{c})}}\frac{v^{a}v^{b}}{f(r_{c})}(\frac{f(r)}{r^{2}})_{c}^{\prime}dx^{a}dx^{b}\nonumber \\
 &  & +\frac{r_{c}^{4}}{2\sqrt{f(r_{c})}}[\frac{r_{c}^{3}}{2f(r_{c})}(\frac{f(r)}{r^{2}})_{c}^{\prime2}-(n-1)(\frac{f(r)}{r^{2}})_{c}^{\prime}-r_{c}(\frac{f(r)}{r^{2}})_{c}^{\prime\prime}]Pdx^{a}dx^{a}\nonumber \\
 &  & -\frac{r_{c}^{2}}{2\sqrt{f(r_{c})}}[1+f(r_{c})\mathcal{F}^{\prime}(r_{c})](\partial_{a}v^{b}+\partial_{b}v^{a})dx^{a}dx^{b}+\mathcal{O}(\epsilon^{3}),\label{eq:perturbed}\end{eqnarray}
after imposing the incompressibility condition (\ref{eq:incompress})
in the $\mathcal{O}(\epsilon^{2})$ part.%
\footnote{Our result above, with $f^{\prime\prime}(r_{c})$ presenting, is slightly
different from (27) in \cite{Cai}, where there is no $f^{\prime\prime}(r_{c})$.
But in the case considered in \cite{Cai} (without electromagnetic
field), we have checked that these two expressions give the same result.%
} The transverse components of the Einstein equations give the conservation
law
\begin{equation}
D^{i}t_{ij}=n^{\mu}T_{\mu j}=F_{ji}J^{i},\qquad J^i\equiv -n_\mu F^{\mu i}\label{eq:conserve}
\end{equation}
of the Brown-York tensor, where $D$ is the
covariant derivative on the cutoff surface, $n$ the unit normal
of the surface, $F_{ji}$ the boundary
electromagnetic field and $J^i$ the boundary current dual to the bulk electromagnetic field (see
e.g. \cite{Strominger}). In our case, the cutoff surface is always flat, so
$D$ is just $\partial$.
For the perturbed metric (\ref{eq:metric}), it can be explicitly checked that\[ n^{\mu}T_{\mu
j}=\mathcal{O}(\epsilon^{3}).\] So the leading order equation of the
index $j=t$ in (\ref{eq:conserve}) is\[
\partial^{i}t_{it}=-\frac{r_{c}^{2}}{16\pi G\sqrt{f(r_{c})}}(\frac{f(r)}{r^{2}})_{c}^{\prime}\partial_{a}v^{a}=F_{ta}J^{a}=0\]
at $\mathcal{O}(\epsilon^{2})$, which is just the incompressibility
condition (\ref{eq:incompress}). The leading order equations of the
index $j=a$ in (\ref{eq:conserve}) are\begin{equation}
\partial^{i}t_{ia}=\frac{r_{c}^{4}}{16\pi Gf(r_{c})\sqrt{f(r_{c})}}\{(\frac{f(r)}{r^{2}})_{c}^{\prime}(\partial_{t}v^{a}+v^{b}\partial_{b}v^{a}+\frac{c}{r_{c}^{2}}\partial_{a}P)-\frac{f(r_{c})}{r_{c}^{2}}[1+f(r_{c})\mathcal{F}^{\prime}(r_{c})]\partial^{2}v^{a}\}=f_{a}\label{eq:N-S}\end{equation}
at $\mathcal{O}(\epsilon^{3})$, where we have
defined\begin{equation}
c\equiv\frac{1}{2}r_{c}^{3}(\frac{f(r)}{r^{2}})_{c}^{\prime}-(n-1)f(r_{c})-r_{c}f(r_{c})(\frac{f(r)}{r^{2}})_{c}^{\prime\prime}/(\frac{f(r)}{r^{2}})_{c}^{\prime}\label{eq:c}\end{equation}
and $f_{a}\equiv F_{ai}J^{i}$ as the external force density (see
e.g. \cite{D-L}).

Now we can read off the viscosity from the Brown-York tensor (\ref{eq:perturbed}).
In fact, the stress-energy tensor of a (relativistic) viscous fluid
is\begin{equation}
t_{ij}=(\rho+p)u_{i}u_{j}+pg_{ij}-2\eta\sigma_{ij}-\zeta\theta(g_{ij}+u_{i}u_{j})\label{eq:viscous}\end{equation}
with $\sigma_{ij}$ the shear and $\theta=\partial_{i}u^{i}$ the
expansion. We are only interested in the above stress-energy tensor
up to $\mathcal{O}(\epsilon^{2})$ under the non-relativistic expansion.
To this order we have $\theta=0$ by the incompressibility condition
(\ref{eq:incompress}), which renders the last term in (\ref{eq:viscous})
to vanish, and so \[
\sigma_{ij}=\frac{1}{2}\mathcal{P}(\partial_{i}u_{j}+\partial_{j}u_{i}),\]
where $\mathcal{P}$ means projection to the $x^{a}$ directions.
Comparing (\ref{eq:viscous}) with (\ref{eq:perturbed}), we have\[
\eta=\frac{1+f(r_{c})\mathcal{F}^{\prime}(r_{c})}{16\pi G}=\frac{1}{16\pi G}\frac{r_{h}^{n-2}}{r_{c}^{n-2}},\]
which together with (\ref{eq:entropy}) gives the cutoff-independent
result\[
\frac{\eta}{s_{c}}=\frac{1}{4\pi}.\]
Under the special rescaling (\ref{eq:special}) (with constant $P$)
of the background configuration (\ref{eq:background}), one finds
that the transformed energy density and pressure become\begin{eqnarray*}
\rho_{s} & = & \rho+\frac{(n-2)r_{c}^{2}}{16\pi G\sqrt{f(r_{c})}}(\frac{f(r)}{r^{2}})_{c}^{\prime}P+\mathcal{O}(\epsilon^{3}),\\
p_{s} & = & p+\frac{r_{c}^{2}}{16\pi
G\sqrt{f(r_{c})}}[\frac{r_{c}^{3}}{2f(r_{c})}(\frac{f(r)}{r^{2}})_{c}^{\prime2}-(n-1)(\frac{f(r)}{r^{2}})_{c}^{\prime}-r_{c}(\frac{f(r)}{r^{2}})_{c}^{\prime\prime}]P+\mathcal{O}(\epsilon^{3}),\end{eqnarray*}
respectively. So the ratio of pressure (or {}``pressure density'')
should be \cite{BMW,Cai}\[
P_{r}=\frac{p_{s}-p}{\rho+p}=\frac{1}{16\pi
G}[\frac{r_{c}^{3}}{2f(r_{c})}(\frac{f(r)}{r^{2}})_{c}^{\prime}-(n-1)-r_{c}(\frac{f(r)}{r^{2}})_{c}^{\prime\prime}/(\frac{f(r)}{r^{2}})_{c}^{\prime}]P=\frac{cP}{f(r_{c})},\]
using (\ref{eq:invariant}) and (\ref{eq:c}). Introducing the
coordinates
\begin{equation}\label{eq:standard}
\tilde{t}=\sqrt{f(r_{c})}t,\qquad\tilde{x}^{a}=r_{c}x^{a},
\end{equation}
under which the induced metric on the cutoff surface becomes
$\eta_{ij}$, one can easily recast the conservation equation
(\ref{eq:N-S}) as the standard incompressible Navier-Stokes equation
\begin{equation}\label{eq:incompress_NS}
\tilde{\partial}_{t}\beta^{a}+\beta^{b}\tilde{\partial}_{b}\beta^{a}+\tilde{\partial}_{a}P_{r}-\nu\tilde{\partial}^{2}\beta^{a}=\tilde{f}_{a},\qquad
\tilde{f}_{a}=\frac{16\pi G\sqrt{f(r_{c})}}{r_{c}^{3}(\frac{f(r)}{r^{2}})_{c}^{\prime}}f_{a}
\end{equation}
with external force density $\tilde{f}_{a}$, where the kinematic viscosity
\begin{equation}
\nu=\frac{\eta}{\rho+p}.
\end{equation}

\section{The Gauss-Bonnet-Maxwell case}

The action of the $n$-dimensional Gauss-Bonnet-Maxwell gravity is
\[I=\frac{1}{16\pi G}\int d^{n}x\sqrt{-g}(R-2\Lambda+\alpha\mathcal {L}_{GB})-\frac{1}{4}\int d^{n}x\sqrt{-g}F_{\mu\nu}F^{\mu\nu},\]
\[\mathcal{L}_{GB}=R^2-4R_{\mu \nu}R^{\mu \nu}+R_{\mu \nu \sigma\tau}R^{\mu \nu \sigma \tau}\]
with $\alpha$ the Gauss-Bonnet coupling constant. The corresponding
equations of motion are
\begin{eqnarray}
G_{\mu\nu}+\Lambda g_{\mu\nu}+\alpha H_{\mu \nu}+8\pi GT_{\mu\nu} = 0,\nonumber \\
\nabla_{\mu}F^{\mu\nu} = 0,\label{eq:GB-M}
\end{eqnarray}
where
\begin{equation}
H_{\mu \nu}=2(R_{\mu \sigma \kappa \tau }R_{\nu }^{\phantom{\nu}
\sigma \kappa \tau }-2R_{\mu \rho \nu \sigma }R^{\rho \sigma
}-2R_{\mu \sigma }R_{\phantom{\sigma}\nu }^{\sigma }+RR_{\mu \nu
})-\frac{1}{2}g_{\mu \nu }\mathcal {L}_{GB}.
\end{equation}

We again take the charged black brane solution \cite{Cai2}
\begin{eqnarray}
ds^{2} & = & -f(r)dt^{2}+2drdt+r^{2}dx^{a}dx^{a},\nonumber \\
f(r)&=&\frac{r^2}{2\tilde{\alpha}}\Bigg(1-\sqrt{1-4\tilde{\alpha}(1-\frac{2m}{r^{n-1}}+\frac{Q^2}{r^{2n-4}})}\Bigg)
\label{eq:background-GB}
\end{eqnarray}
as our background, where we have defined
$\tilde{\alpha}=(n-3)(n-4)\alpha$ for convenience. The
electromagnetic field is of the same form as in the Einstein-Maxwell
case.

It turns out that the non-relativistic long-wavelength expansion for
the perturbed configurations in the Einstein-Maxwell case, i.e. the
perturbed metric (\ref{eq:metric}) and electromagnetic field
(\ref{eq:em_field}) up to $\mathcal{O}(\epsilon^2)$, as well as the
form (\ref{eq:correction}) of the correction term to the metric
(\ref{eq:metric}), can also be used in the Gauss-Bonnet-Maxwell
case. Again, the Maxwell equations are automatically satisfied at
$\mathcal{O}(1)$ and $\mathcal{O}(\epsilon)$, while at
$\mathcal{O}(\epsilon^{2})$ lead to two equations
\begin{equation}
\partial_{a}v^{a} = 0,\label{eq:incompress-GB}\qquad
\mathcal{F}^{\prime}(r)\partial_{a}v^{a} = 0,
\end{equation}
which can be solved altogether by the incompressibility condition.
Then from the Einstein equations at $\mathcal{O}(\epsilon^{2})$
there is again only one requirement
for $\mathcal{F}(r)$ to satisfy, which can be
solved as\footnote{We have verified this solution for $5\le n\le10$.}
\[\mathcal{F}^{\prime}(r)=-\Big(1-\frac{C}{r^2\big(r^{n-4}-2(n-4)\alpha[r^{n-5}f(r)]^{\prime}\big)}\Big)\frac{1}{f(r)}.\]
Here the integration constant $C$ can be determined as
$C=r_h^{n-2}-2(n-4)\alpha r_h^{n-3}f^{\prime}(r_h)$ with $r_{h}$ the
horizon radius by the regularity condition of the perturbed metric
at the horizon.

The Brown-York tensor on the cutoff surface $r=r_c$ in this case is
\cite{B-R,Cai,ABD}
\begin{equation}
t_{ij}=\frac{1}{8 \pi G}[Kg _{ij}-K_{ij}-2\alpha
(3J_{ij}-Jg_{ij})-\mathcal{C}g _{ij}]\label{eq:B-Y-GB}
\end{equation}
with
\begin{equation}
J_{ij}=\frac{1}{3}(2KK_{ik}K_{j}^{k}+K_{kl}K^{kl}K_{ij}-2K_{ik}K^{kl}K_{lj}-K^{2}K_{ij}),
\end{equation}
where we have omitted terms that do not contribute for this flat
cutoff surface with induced metric (\ref{eq:cutoff}). It can be
explicitly worked out for the background metric
(\ref{eq:background-GB}) that
\begin{eqnarray}
t_{ij}dx^{i}dx^{j} &=& \frac{1}{8\pi G}\Big[\sqrt{f(r_{c})}\frac{(n-2)f(r_{c})}{r_{c}}\Big(\frac{2\tilde{\alpha} f(r_{c})}{3r^2_{c}}-1\Big)dt^{2}\nonumber \\
&&+\frac{r_{c}^{2}}{\sqrt{f(r_{c})}}\bigg(\frac{f^{\prime}(r_{c})}{2}+\frac{(n-3)f(r_{c})}{r_{c}}-\tilde{\alpha} f(r_c)\Big(\frac{2(n-5)f(r_c)}{3r_c^3}+\frac{f^{\prime}(r_c)}{r_c^2}\Big)\bigg)dx^{a}dx^{a}\nonumber \\
&&-\mathcal{C}ds_{c}^{2}\Big].\label{eq:B-Y-GB c}
\end{eqnarray}
Comparing (\ref{eq:B-Y-GB c}) and (\ref{eq:fluid}), we find
\begin{eqnarray}
\rho+p &=&\frac{1}{8\pi
G\sqrt{f(r_{c})}}[(n-1)\frac{m}{r_{c}^{n-2}}-(n-2)\frac{Q^{2}}{r_{c}^{2n-5}}]
.\label{eq:invariant-GB}
\end{eqnarray}
The entropy density $s_{c}$ and local Hawking temperature
$T_{c}$ on the cutoff surface are of the same form (\ref{eq:entropy}) and (\ref{eq:temperature}) as in the
Einstein-Maxwell case, which again leads to the thermodynamic
relation (\ref{eq:Gibbs}) with charge density (\ref{eq:charge}) and
chemical potential (\ref{eq:chemical}).

For the perturbed metric (\ref{eq:metric}), the Brown-York tensor on
the cutoff surface can be worked out as
\begin{eqnarray}
8\pi G t_{ij}dx^{i}dx^{j} &=& \sqrt{f(r_{c})}\frac{(n-2)f(r_{c})}{r_{c}}\Big(\frac{2\tilde{\alpha} f(r_{c})}{3r^2_{c}}-1\Big)dt^{2}-\mathcal{C}ds_{c}^{2}\nonumber \\
&&+\frac{r_{c}^{2}}{\sqrt{f(r_{c})}}\bigg(\frac{f^{\prime}(r_{c})}{2}+\frac{(n-3)f(r_{c})}{r_{c}}-\tilde{\alpha} f(r_c)\Big(\frac{2(n-5)f(r_{c})}{3r_c^3} +\frac{f^{\prime}(r_c)}{r_c^2}\Big)\bigg)dx^{a}dx^{a}\nonumber \\
&&+\bigg(\frac{2\tilde{\alpha} f(r_c)}{r_c^2}-1\bigg)\frac{r_{c}^{4}}{\sqrt{f(r_{c})}}(\frac{f(r)}{r^{2}})_{c}^{\prime}v^{a}dx^{a}dt\nonumber \\
&&+\bigg(1-\frac{2\tilde{\alpha}
f(r_c)}{r_c^2}\bigg)\bigg((n-2)f(r_{c})P+r_{c}^{2}v^{2}\bigg)\frac{r_{c}^{2}}{2\sqrt{f(r_{c})}}(\frac{f(r)}{r^{2}})_{c}^{\prime}dt^{2}
\nonumber \\
&&+\bigg(1-\frac{2\tilde{\alpha}
f(r_c)}{r_c^2}\bigg)\frac{r_{c}^{6}}{2\sqrt{f(r_{c})}}\frac{v^{a}v^{b}}{f(r_{c})}(\frac{f(r)}{r^{2}})_{c}^{\prime}dx^{a}dx^{b}
\nonumber \\
&&+\bigg(1-\frac{2\tilde{\alpha}
f(r_c)}{r_c^2}\bigg)\frac{r_{c}^{4}}{2\sqrt{f(r_{c})}}
\bigg(\frac{r_{c}^{3}\bigg(r_c^2+2\tilde{\alpha} f(r_c)\bigg)}{2f(r_{c})\bigg(r_c^2-2\tilde{\alpha} f(r_c)\bigg)}(\frac{f(r)}{r^{2}})_{c}^{\prime2}\nonumber \\
&&-(n-1)(\frac{f(r)}{r^{2}})_{c}^{\prime}-r_{c}(\frac{f(r)}{r^{2}})_{c}^{\prime\prime}\bigg)Pdx^{a}dx^{a}
\nonumber \\
&&-\bigg(1-2\tilde{\alpha}\big(\frac{f(r_c)}{r_c^2}+\frac{r_c}{n-3}(\frac{f(r)}{r^{2}})_{c}^{\prime}\big)\bigg)
\frac{r_{c}^{2}[1+f(r_{c})\mathcal{F}^{\prime}(r_{c})]}{2\sqrt{f(r_{c})}}
(\partial_{a}v^{b}+\partial_{b}v^{a})dx^{a}dx^{b}\nonumber \\
&&+\mathcal{O}(\epsilon^{3}), \label{eq:perturbed-GB}
\end{eqnarray}
after imposing the incompressibility condition
(\ref{eq:incompress-GB}) in the $\mathcal{O}(\epsilon^{2})$
part.\footnote{Similar to the Einstein-Maxwell case, our result
above is different from (57) in \cite{Cai}, but we have checked that
these two expressions give the same result when turning off the
electromagnetic field.} So the leading order equation of the index
$j=t$ in (\ref{eq:conserve}) is
\[\partial^{i}t_{it}=(\frac{2\tilde{\alpha} f(r_c)}{r_c^2}-1)\frac{r_{c}^{2}}{16\pi G\sqrt{f(r_{c})}}(\frac{f(r)}{r^{2}})_{c}^{\prime}\partial_{a}v^{a}=n^{\mu}T_{\mu t}=0\]
at $\mathcal{O}(\epsilon^{2})$, which is just the incompressibility
condition (\ref{eq:incompress-GB}). The leading order equations of
the index $j=a$ in (\ref{eq:conserve}) are
\begin{eqnarray}
\partial^{i}t_{ia}&=&\frac{r_{c}^{4}}{16\pi G f(r_{c})\sqrt{f(r_{c})}}\{(1-\frac{2\tilde{\alpha}f(r_c)}{r_c^2})(\frac{f(r)}{r^{2}})_{c}^{\prime}(\partial_{t}v^{a}
+v^{b}\partial_{b}v^{a}+\frac{c}{r_c^2}\partial_{a}P)\nonumber \\
&&-\frac{f(r_{c})}{r_{c}^{2}}\bigg(1-2\tilde{\alpha}\big(\frac{f(r_c)}{r_c^2}+\frac{r_c}{n-3}(\frac{f(r)}{r^{2}})_{c}^{\prime}\big)\bigg)
[1+f(r_{c})\mathcal{F}^{\prime}(r_{c})]\partial^{2}v^{a}\}=f_{a}
\label{eq:N-S-GB}
\end{eqnarray}
at $\mathcal{O}(\epsilon^{3})$, where we have defined
\begin{equation}
c\equiv\frac{r_c^3(r_c^2+2\tilde{\alpha}f(r_c))}{2(r_c^2-2\tilde{\alpha}f(r_c))}(\frac{f(r)}{r^{2}})_{c}^{\prime}
-(n-1)f(r_{c})-r_cf(r_{c})(\frac{f(r)}{r^{2}})_{c}^{\prime\prime}/(\frac{f(r)}{r^{2}})_{c}^{\prime}
\label{eq:c-GB}
\end{equation}
and $f_{a}\equiv n^{\mu}T_{\mu a}=F_{ai}J^{i}$ as the external force
density.

Now we can read off the viscosity from the Brown-York tensor
(\ref{eq:perturbed-GB}). Comparing (\ref{eq:viscous}) with
(\ref{eq:perturbed-GB}), we have
\begin{eqnarray}
\eta&=&\frac{1}{16\pi
G}\frac{r_{h}^{n-2}}{r_{c}^{n-2}}\bigg(1-2\tilde{\alpha}\big(\frac{n-1}{n-3}-q_h^2\big)\bigg)
\end{eqnarray}
with the charge density of the black brane
$q_h=\frac{Q}{r_h^{n-2}}$, which together with (\ref{eq:entropy})
gives the cutoff-independent result
\begin{equation}
\frac{\eta}{s_{c}}=\frac{1}{4\pi}\bigg(1-2\tilde{\alpha}\big(\frac{n-1}{n-3}-q_h^2\big)\bigg)=\frac{1}{4\pi}\bigg(1-2(n-4){\alpha}[n-1-(n-3) q_h^2]\bigg).
\end{equation}
This ratio agrees with the known result for infinite boundary
$r_c\to\infty$ \cite{Cai3,GMSST,G-S,HSZ}. Under the special
rescaling (\ref{eq:special}) (with constant $P$) of the background
configuration (\ref{eq:background-GB}), one finds that the
transformed energy density and pressure become
\begin{eqnarray*}
\rho_{s} & = & \rho+\frac{(n-2)r_{c}^{2}}{16\pi G\sqrt{f(r_{c})}}(1-\frac{2\tilde{\alpha}f(r_c)}{r_c^2})(\frac{f(r)}{r^{2}})_{c}^{\prime}P+\mathcal{O}(\epsilon^{3}),\\
p_{s} & = & p+\frac{r_{c}^{2}}{16\pi
G\sqrt{f(r_{c})}}(1-\frac{2\tilde{\alpha}f(r_c)}{r_c^2})
[\frac{r_{c}^{3}(r_c^2+2\tilde{\alpha}f(r_c))}{2f(r_{c})(r_c^2-2\tilde{\alpha}f(r_c))}(\frac{f(r)}{r^{2}})_{c}^{\prime2} \\
&&-(n-1)(\frac{f(r)}{r^{2}})_{c}^{\prime}-r_{c}(\frac{f(r)}{r^{2}})_{c}^{\prime\prime}]P+\mathcal{O}(\epsilon^{3}),
\end{eqnarray*}
respectively. So the ratio of pressure should be\[
P_{r}=\frac{p_{s}-p}{\rho+p}=[\frac{r_{c}^{3}(r_c^2+2\tilde{\alpha}f(r_c))}{2f(r_{c})(r_c^2-2\tilde{\alpha}f(r_c))}(\frac{f(r)}{r^{2}})_{c}^{\prime}-(n-1)
-r_{c}(\frac{f(r)}{r^{2}})_{c}^{\prime\prime}/(\frac{f(r)}{r^{2}})_{c}^{\prime}]P=\frac{cP}{f(r_{c})},\]
using (\ref{eq:invariant-GB}) and (\ref{eq:c-GB}). Introducing the
standard coordinates (\ref{eq:standard}), one can again recast the
conservation equation (\ref{eq:N-S-GB}) as the standard
incompressible Navier-Stokes equation (\ref{eq:incompress_NS}) with
external force density.

\section{Concluding remarks}

Under the non-relativistic long-wavelength expansion, we have solved
up to second order of the expansion parameter the bulk equations of
motion for Dirichlet-like boundary conditions at an arbitrary cutoff
surface outside the horizon in the charged AdS black brane
space-times in both the Einstein-Maxwell and Gauss-Bonnet-Maxwell
theories{, without considering the independent electromagnetic DoF}. The incompressible Navier-Stokes equation with external
force density, as well as a cutoff-independent viscosity to entropy
density ratio $\frac{\eta}{s}$, for the dual fluid on the cutoff
surface has been obtained in both theories, {while in
the Gauss-Bonnet-Maxwell case the ratio
$\frac{\eta}{s}=\frac{1}{4\pi}\{1-2(n-4){\alpha}[n-1-(n-3) q_h^2]\}$ depends on the charge density of the black brane.}

The framework used in our Letter seems to work well in all known
cases of intrinsically flat cutoff surfaces, {but for general curved cutoff
surfaces it can no longer be used. For some special curved cutoff surface, there have been some discussions \cite{B-S}.
Moreover,} it is worthy to point out that very recently an
alternative method has been introduced in \cite{L-S} by imposing
Petrov type I condition on the cutoff surface. It turns out that
imposing this boundary condition is equivalent to imposing
regularity condition on the horizon at least in the near horizon
limit such that the Navier-Stokes equation can be
derived in a much simpler way. Following this approach the
generalized framework applicable to the spatially curved space-time
has been presented in \cite{HLPTW}, and we expect that this strategy
can be applied to our current work in future.

\begin{acknowledgments}
This work is partly supported by the National Natural Science
Foundation of China (Grant Nos. 10731080, 10875057 and {11075206}).
Y.~Ling also acknowledges the support by Jiangxi Young Scientists
(JingGang Star) program, 555 talent project of Jiangxi Province and
the Program for Innovative Research Team of Nanchang University.
\end{acknowledgments}

\end{document}